\documentclass{ws-procs9x6}
\usepackage{epsfig}

\begin{document}

\title{Cascades from $\nu_e$ above $10^{20}$ eV}

\author{Spencer R. Klein}

\address{Nuclear Science Division, Lawrence Berkeley National Laboratory, \\
Berkeley, CA, 94720, USA\\E-mail: srklein@lbl.gov}

\maketitle

\abstracts{At very high energies, the Landau-Pomeranchuk-Migdal effect
reduces the cross sections for electron bremsstrahlung and photon
$e^+e^-$ pair production.  The fractional electron energy loss and
pair production cross sections drop as the energy increases.  In
contrast, the cross sections for photonuclear interactions grow with
energy.  In solids and liquids, at energies above $10^{20}$ eV,
photonuclear reactions dominate, and showers that originate as photons
or electrons quickly become hadronic showers.  These
electron-initiated hadronic showers are much shorter (due to the
absence of the LPM effect), but wider than purely electromagnetic
showers would be.  This change in shape alters the spectrum of the
electromagnetic and acoustic radiation emitted from the shower.  These
alterations have important implications for existing and planned
searches for radiation from $\nu_e$ induced showers above $10^{20}$
eV, and some existing limits should be reevaluated.}

\section{Introduction}

Although ultra-high energy (UHE) cosmic rays have been studied for
many years, their origin is still a mystery.  Many cosmic-ray models
predict significant fluxes of astrophysical neutrinos with energies
above $10^{20}$ eV.  Proposed models consider topological defects,
superheavy relics of the big bang\cite{kusenko} and UHE neutrinos as
cosmic rays\cite{nucosmicrays}.  Conventional approaches predict that
the GZK mechanism\cite{GZK} and gamma ray bursts\cite{GRB} produce
neutrinos with energies above $10^{20}$ eV.

Several groups have searched for radio or acoustic radiation from
electromagnetic cascades produced by interacting $\nu_e$ and have
reported upper limits on the cosmic flux of $\nu_e$ (here, $\nu_e$
includes $\overline{\nu_e}$) at energies of $10^{20}$ to $10^{25}$ eV.
These searches probe enormous volumes to reach interesting upper
limits.  These limits depend on a good understanding of the cascades
that are produced in $\nu_e$ interactions.

\section{Radio and Acoustic Waves from Showers}

Most of the searches have involved radio waves.  The Glue
collaboration searched for $\approx 2.3$ GHz radio waves from the moon using
two radio-telescopes\cite{glue}.  FORTE used satellite-based receivers
to search for 30-300 MHz radiation from the Greenland ice
pack\cite{forte}.  Both ANITA\cite{ANITA}, a balloon-based detector,
and RICE\cite{rice}, a surface antenna array will search for radio waves
from $\nu_e$ cascades in Antartic ice. The SalSA collaboration plans
to search from radio emission from underground salt domes\cite{salsa}.

Translating these search results into a $\nu_e$ limit requires a model
of the electromagnetic radiation produced by the shower.  This
radiation has been evaluated using a 3-d Monte Carlo shower
simulations\cite{zhs}.  The calculations add the electromagnetic
fields from each particle in the shower.  This is a computationally
demanding process which is only practical for relatively low energy
showers, below 1 PeV.  At higher energies, extrapolations are
used\cite{zasEM,zascalc}.

The radiation is large when the electromagnetic fields from the
different particles add coherently\cite{ask}, and the radiated
energy is proportional to the square of the shower energy.  This
happens when the radio wavelength is larger than the transverse spread
of the shower, seen along the direction of propagation.  The positive
and negative charges cancel but, because of positron annihilation, the
overall shower contains about 20\% more $e^-$ than $e^+$.  This
electron excess produces the coherent radiation.  When the wavelength
is short compared to the lateral size, phase coherence is lost.
Radiation from the individual particles adds incoherently, producing a
much smaller signal.  The degree of coherence depends on the width of
the shower; current measurements are largely, but not completely in
the coherent domain, so the radiation is sensitive to the transverse
shower spreading.

The SAUND collaboration has searched for acoustic radiation from
$\nu_e$ induced showers, using data from a set of U.S. Navy
hydrophones in Bermuda\cite{saund}.  The acoustic pulse is generated
when the electromagnetic cascade heats the water, causing it to expand
rapidly around the cascade.  The frequencies are low (in the audio
range) because the speed of sound is so much lower than that of light.
The pulse strength is proportional to the energy deposition.  The
acoustic frequencies are subject to comparable coherence conditions as
the radio studies.  The frequency spectrum of the radiation again
depends on the shower width.

None of the calculations used in these studies considers the effect of
photonuclear interactions on electromagnetic showers.  Here, we show
that photonuclear interactions significantly alter $\nu_e$ induced
showers, and discuss how these interactions affect the shower shape
and emitted radiation.

\section{Electromagnetic Interactions at High Energies}

In high-energy $\nu_e$ interactions, the produced electrons receive an
average of 80\% of the $\nu_e$ energy.  The remainder is transferred
to the target nucleus, producing a hadronic shower.

Electrons with energies $E> 100$ MeV lose energy largely via
bremsstrahlung.  At somewhat higher energies, electrons lose their
energy over a distance scale of order $X_0$, the radiation length.
For ice, $X_0 = 36.1$g/cm$^2$.  The density of ice depends on its
composition (mostly air content).  In Antartica, the ice is covered by
a layer of compressed snow. For simplicity here, we use a uniform
medium with a density of 1 g/cm$^2$ (as for water), so $X_0 =$ 36.1
cm.  Almost all ice is within 10\% of this value; snow has a somewhat
lower density.

At very high energies, the Landau-Pomeranchuk-Migdal (LPM) effect
suppresses the bremsstrahlung of low-energy photons, increasing the
distance scale.  Radiation of photons with energy $k$ from electrons
with energy $E$ is suppressed when \cite{lp,lpmreview}
\begin{equation}
k < \frac{E(E-k)}{E_{LPM}}
\end{equation}
where $E_{LPM}$ is a material dependent constant,
\begin{equation}
E_{LPM} = \frac{m^4X_0}{E_s^2} \approx 7.7\ {\rm TeV/cm}\cdot X_0.
\end{equation}
Here $m$ is the mass of the electron, $E_s=m\sqrt{4\pi/\alpha} = 21.2$
MeV, and $\alpha\approx 1/137$ is the fine structure constant.  When
$E>E_{LPM}$, the effective radiation length $X \approx
X_0 \sqrt{E/E_{LPM}}$.

\begin{figure}[tb]
\centerline{\psfig{figure=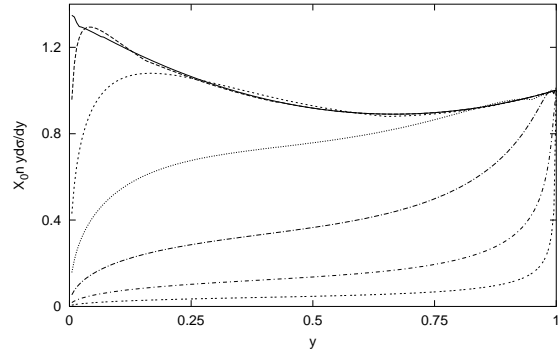,height=3.0 in,angle=270,clip=}}
\caption{The energy weighted differential cross section per radiation
length for bremsstrahlung in water as a function of $y=k/E$ for
electrons with energies of 640 GeV (top curve), 6.4 TeV, 64 TeV, 640
TeV, 6.4 PeV, 64 PeV and 640 PeV (bottom curve).  These curves apply
for other materials for electron energies of 0.0023, 0.023, 0.23, 2.3,
23, 230 and 2300 times $E_{LPM}$.}
\label{fig:bremscaling}
\end{figure}

\begin{figure}[tb]
\centerline{\psfig{figure=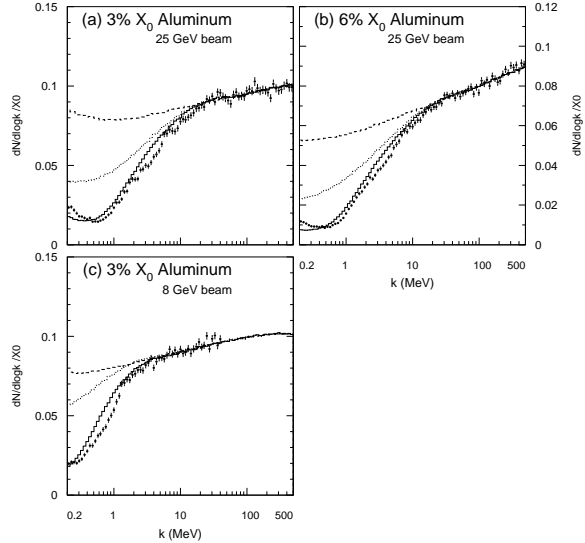,height=3.0in,clip=}}
\caption{Data (points) from 8 and 25 GeV electrons passing through
aluminum targets, from the SLAC E-146 collaboration, compared with
calculations based on the Bethe-Heitler (dashed histogram), and
Migdals LPM suppression (dot-dashed histogram), and on Migdals
calculations with LPM and dielectric suppression (solid histogram).
The data is binned logarithmically in photon energy, so the
Bethe-Heitler $1/k$ spectrum is roughly flat.}
\label{fig:e146}
\end{figure}

Most shower studies calculate LPM suppression using Migdals 1956
calculation of this suppression, albeit with some numerical
simplifications\cite{stanev}.  The degree of suppression depends on a
variable $s$
\begin{equation}
s= \sqrt{\frac{E_{LPM}k}{8E(E-k)\xi(s)}},
\end{equation}
where $1< \xi(s) <2$ increases slowly with $s$; $s\rightarrow\infty$
corresponds to no suppression, while $s\rightarrow 0$ gives strong LPM
suppression.  Figure \ref{fig:bremscaling} shows the differential
bremsstrahlung cross sections for different electron energies.  Figure
\ref{fig:e146} compares Migdals calculations with data from SLAC
experiment E-146.  The figure shows the photon spectrum from 8 and 25
GeV electron beams passing through 3\% and 6\% $X_0$ aluminum
targets\cite{e146prl}. When $k/E < 10^{-4}$, an additional effect,
dielectric suppression, further suppresses
bremsstrahlung\cite{e146diel}. At the same time transition radiation
from the electron entering and exiting the target increases the photon
flux.  An experiment at CERN has observed the increase in effective
radiation length in bremsstrahlung from 149 to 287 GeV
electrons\cite{cernlpm}.

\begin{figure}[tb]
\centerline{\psfig{figure=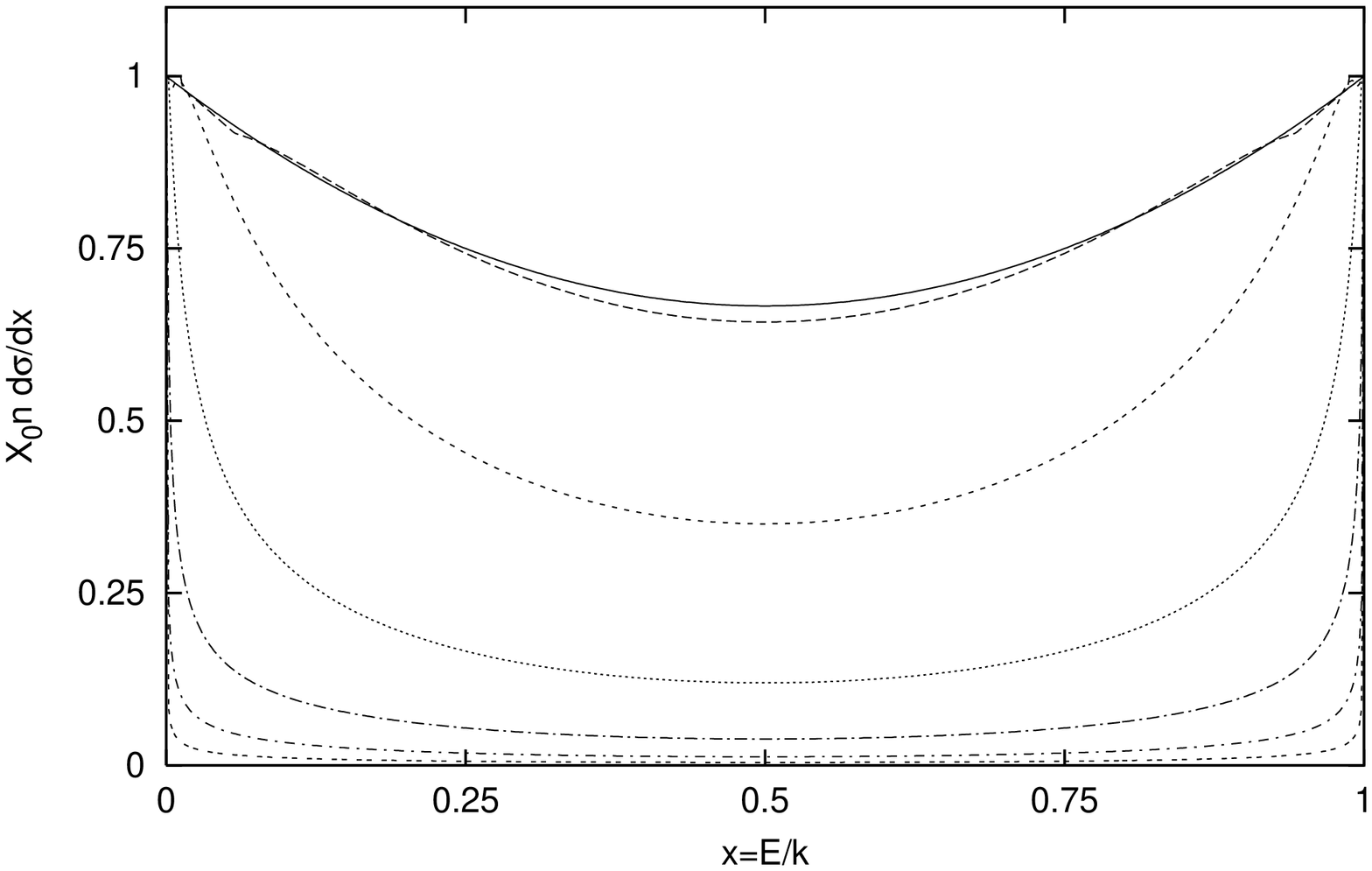,height=3.0in,angle=270,clip=}}
\caption{The differential cross section for pair production in water, as
a function of $x=E/k$ for photons with energies of 64 TeV (top curve),
640 TeV, 6.4 PeV, 64 PeV, 6400 PeV, 6.4 EeV and 64 EeV (bottom curve).
These curves apply for other materials for photon energies of 0.23,
2.3, 23, 230, 2300, 23,000 and 230,000 $E_{LPM}$.}
\label{fig:photonscaling}
\end{figure}

The cross section for pair production may be similarly reduced; when
the photon energy $k$ is greater than $E_{LPM}$, the pair production
cross section is reduced.  Figure \ref{fig:photonscaling} shows the
differential cross section for different photon energies. For a given
photon energy, symmetric pairs are suppressed the most.

\begin{figure}[tb]
\centerline{\epsfig{figure=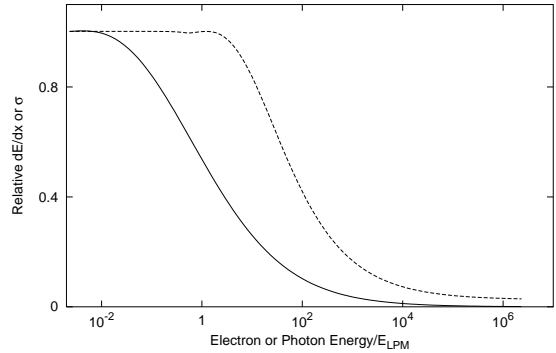,height=3.0in,angle=270,clip=}}
\caption{The electron energy loss ($dE/dx$) for electron
bremsstrahlung (solid line) and the photon pair conversion cross
section (solid line), relative to the Bethe-Heitler predictions ({\it
i.e.}  with no LPM suppression) as a function of $E/E_{LPM}$.}
\label{fig:dedx}
\end{figure}

The reduction in bremsstrahlung cross section corresponds to a
reduction in energy loss ($dE/dx$) by the electron.
Fig. \ref{fig:dedx} shows the reduction in bremsstrahlung $dE/dx$ as a
function of incident electron energy.  Also shown is the reduction in
pair production cross section, as a function of photon energy.
Photons are affected by LPM suppression at much higher energies than
electrons.

\section{Photonuclear Interactions}

In contrast to pair production, for $k> 10$ TeV, the cross section for
photonuclear interactions increases with energy\cite{ralph}.  The
dominant contribution to the photonuclear cross section (from `soft'
interactions) can be described in terms of photon-Pomeron
interactions.  The cross section for photon-proton interactions rises
with the $\gamma p$ center of mass energy $W$ as $W^{0.16}$.  At very
high energies, the photon may also interact directly with a quark in
the target, $\gamma q\rightarrow gq$.  Data from HERA on $\gamma p$
interactions extends up to $W=200$ GeV\cite{pdg}, equivalent to 20 TeV
photons striking stationary protons.  Direct $\gamma q$ interactions
have not clearly been observed at HERA, so predictions about this
process have significant uncertainties.

For oxygen, a Glauber calculation accounts for interactions with
multiple nucleons (shadowing).  This moderates the $W$-dependence of
the cross section.  Since oxygen has only 16 nucleons, the number of
multiple interactions is fairly small.  Here, we assume that the
effect of shadowing in oxygen and direct photon interactions in oxygen
and hydrogen cancel each other out, so the photon-H$_2$O cross section
follows the Pomeron trajectory, $\sigma\approx W^{0.16}$.  This leads
to a lower total cross section than in the original model\cite{ralph},
and a slightly higher crossover energy than is given in
Ref. \cite{pcreview}. For water, this is a reasonable approximation,
but, for heavy nuclei, it may somewhat underestimate the shadowing and
overestimate the cross-section.

\begin{figure}[tb]
\center{
\epsfig{figure=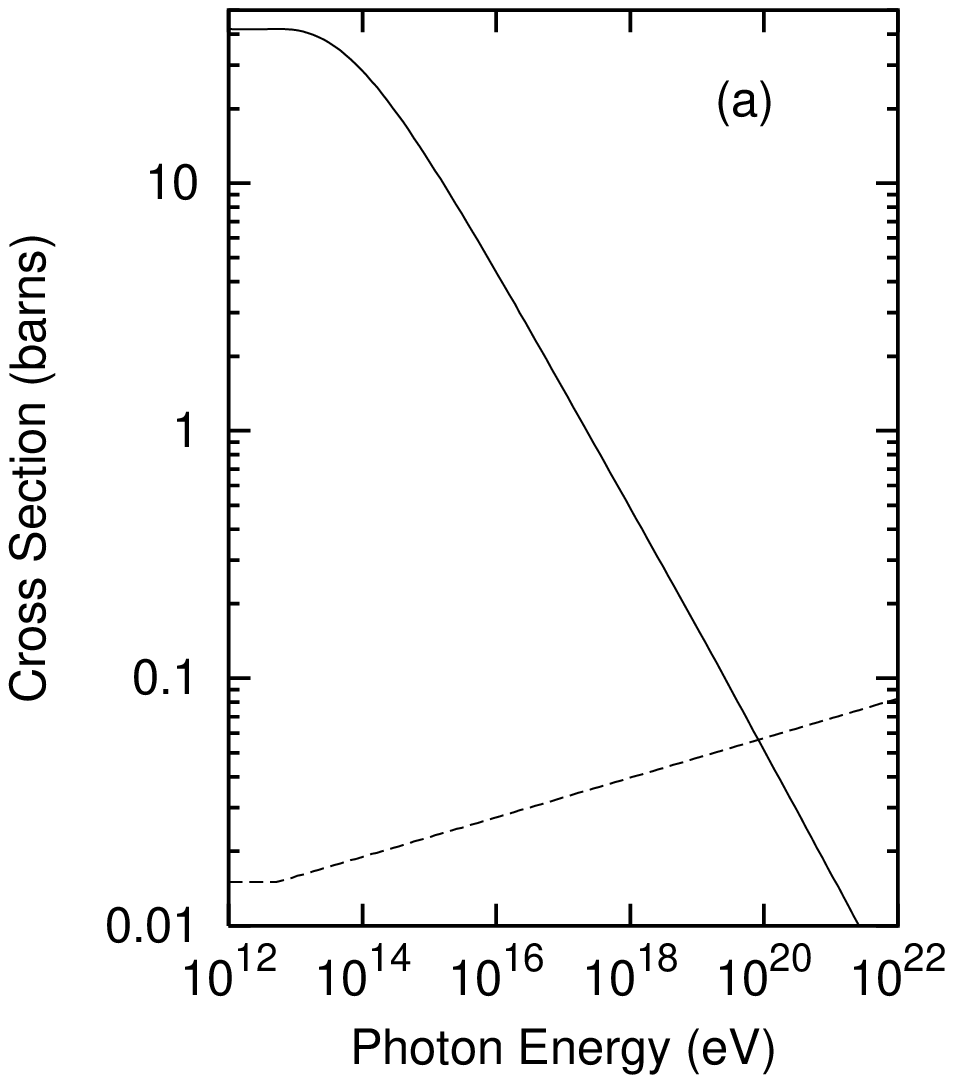,height=2.2in,angle=270,clip=}
\epsfig{figure=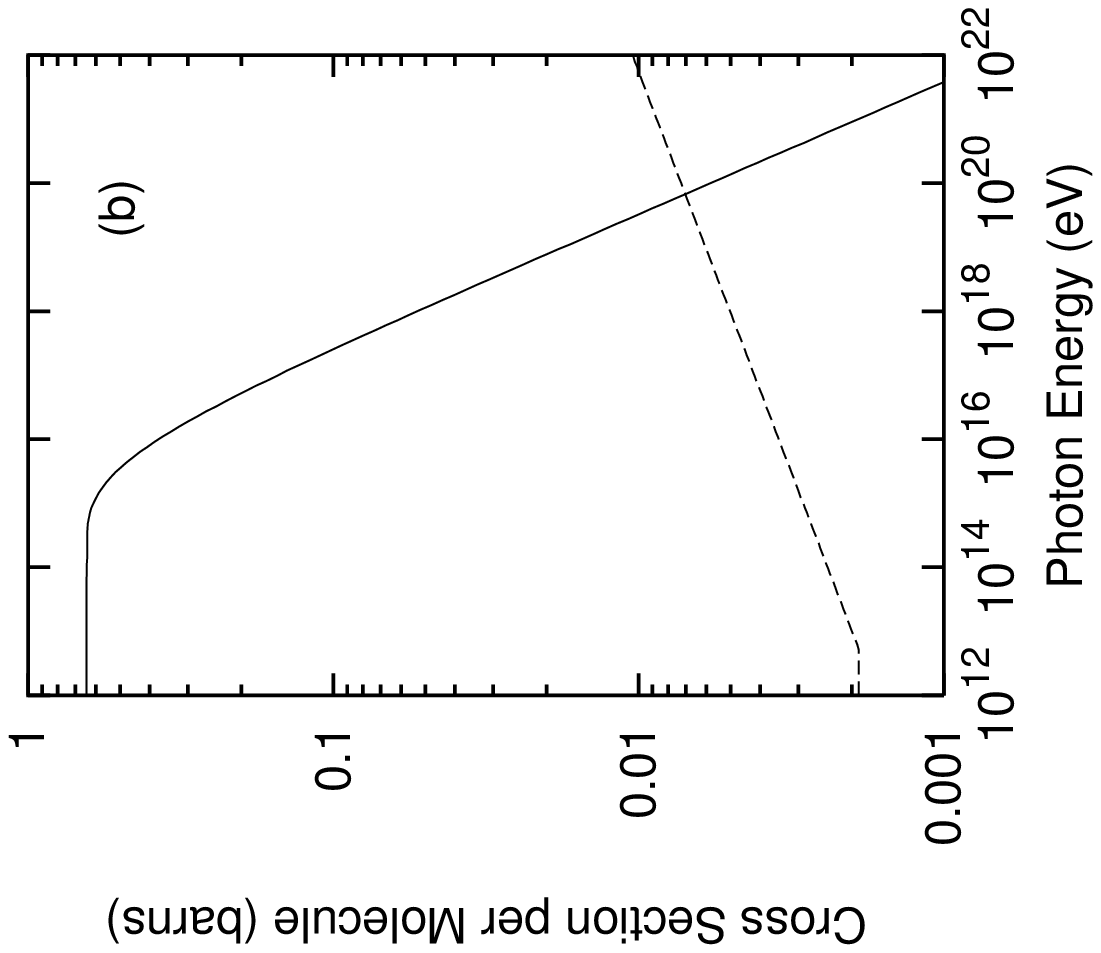,height=2.2in,angle=270,clip=}}
\label{fig:compare}
\caption{Comparison between the pair production and photonuclear
interaction cross sections in (a) lead and (b) water.  Above
an energy of $10^{20}$ eV, the photonuclear cross section
is larger than for pair production.}
\end{figure}

Figure \ref{fig:compare} compares these photonuclear cross sections
with the pair production cross section in lead and water.  In both
materials, photonuclear interactions dominate for $k>10^{20}$ eV.  The
cross-over energy is similar for diverse solids and liquids because
the increase in $\sigma_{\gamma p}/\sigma_{ee}$ for heavier nuclei is
cancelled out by the decrease in $E_{LPM}$ as $X_0$ drops.  Lunar soil
has a density of 1.7 g/cm$^2$\cite{glue}, and should have a similar
crossover energy.  In gasses, because of the reduced density, but
similar atomic number, the crossover point occurs at much higher
energies (about $5\times10^{22}$ eV in air at sea level).

\section{Shower Length}

A $\nu_e$ interaction produces a high-energy electron plus a hadronic
shower from the struck hadron.  Because of the LPM effect, an
energetic electron will travel a long distance before losing its
energy.  Figure \ref{fig:dedx} shows that the electron interaction
distance can be approximated
\begin{equation}
X_e(E) = X_0\sqrt{\frac{2E}{E_{LPM}}}
\end{equation}
for $E>E_{LPM}$.  Here, we neglect electro-nuclear interactions and
direct pair production. At sufficiently high energies, these processes
will be the dominant energy loss mechanisms, and electrons will act
like muons\cite{mmc}.  However, these ultra-high energy electron
interactions have not yet been studied in detail, and so we neglect
them here.

\begin{figure}[tb]
\centerline{\psfig{figure=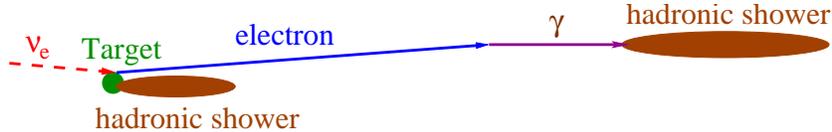,height=0.65in}}
\caption{Diagram of UHE $\nu_e$ shower development with photonuclear
interactions (not to scale).}
\label{fig:schema}
\end{figure}

When the LPM effect is strong ({\it i.e.} above $10^{20}$ eV), the
electron transfers most of its energy to a single photon.  For photon
energy $k\gg E_{LPM}$, the photon pair production distance is
\begin{equation}
X_\gamma(k) = X_0\sqrt{\frac{k}{50E_{LPM}}},
\end{equation}
10 times shorter than $X_e$ at the same energy. When the pair
production cross section falls below the photonuclear cross section,
the photon will usually interact hadronically.  The hadronic
interaction length is $1/\sigma\rho$, where $\sigma$ is the
photonuclear cross section per molecule (from Fig \ref{fig:compare}),
and $\rho$ is the target density.  In water, $\rho=3.3\times10^{22}$
molecules/cm$^3$, and, at $10^{20}$ eV, the hadronic interaction
length in is about 43 m.  Figure \ref{fig:schema} shows schematically
how showers evolve with photonuclear interactions.

We will use a simple model to compare the development of $\nu_e$
cascades with and without photonuclear interactions.  Although
inaccurate, these models are useful for comparison.  Purely
electromagnetic showers evolve via bremsstrahlung and pair production,
with each succeeding shower generation containing twice as many
particles as the last, each with half of the energy.  Shower
development continues until the average particle energy drops below
the critical energy, $E_c$.  Below this energy, Compton scattering
becomes more important than bremsstrahlung and the shower quickly
dissipates its energy into the medium.  This occurs after
$N=\ln_2(E/E_c)$ generations.  In water, $E_c=126$ MeV, so for
$E=10^{20}$ eV, $N=40$.  Without the LPM effect, a generation occurs
in 1 $X_0$, which is 36.1 cm in water.  The shower length is
$40X_0\approx 15$ m.  Some authors\cite{zhs,heitler} have used
$X_0\ln{2}$ for the generation length; this leads to showers that are
30\% shorter than are given here.

The LPM effect does not affect the number of generations. However, the
effective radiation length increases rapidly.  When the LPM effect is
large, the effective radiation length for a generation with average
particle energy $E_g$ is roughly
\begin{equation}
X = X_0 \sqrt{\frac{E_g}{5E_{LPM}}}.
\end{equation}
This length is the geometric average of $X_e(E)$ and
$X_\gamma(k)$ For each successive generation, $X$ decreases by
$1/\sqrt{2}$.

\begin{figure}[tb]
\center{\epsfig{figure=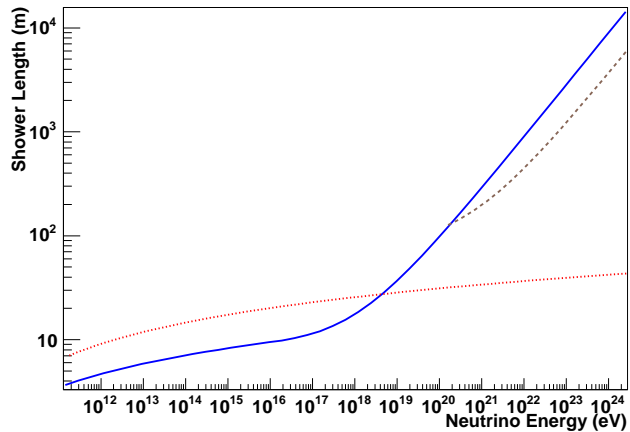,height=2.5in,clip=}}
\caption{Shower length as a function of neutrino energy.  The solid
blue line is for a purely electromagnetic shower, while the dashed
brown line is for a hybrid EM/hadronic shower.  The dotted red line is
for a purely hadronic shower.  The difference between the hybrid and
purely hadronic showers is that the hybrid includes the long electron
path before bremsstrahlung, and the photon path before it interacts
hadronically.  For the hybrid and electromagnetic showers, the
electron is assumed to take 80\% of the $\nu_e$ energy.}
\label{fig:length}
\end{figure}

Figure \ref{fig:length} shows the $\nu_e$ shower length as a function
of energy.  For purely electromagnetic showers, above $10^{18}$ eV,
the LPM effect reduces the cross sections and the length increases
rapidly. This length increase has been experimentally
observed\cite{kasahara}. The LPM effect also increases the
shower-to-shower variation\cite{myMarylandtalk}, complicating
measurements.  

Alvarez-Muniz and Zas also studied the length of electromagnetic
showers\cite{zasEM}.  They defined the length as the distance over
which the shower has more than a given fraction (10\%, 50\% or 70\%)
of the maximum number of particles.  Above $10^{16}$ eV, where the LPM
effect is significant, their length scales as $E^{1/3}$.  For the 70\%
fraction, their lengths are slightly larger than are given here: 6 m
vs. 4.8 m at 1 TeV, and 37 m vs. 20 m at $10^{18}$ eV.  With a
slightly lower containment fraction, the curves would probably agree
fairly well.

For hadronic showers, there are no simple parameterizations.  Because
of the higher final state multiplicity (compared to $e^+e^-$ pairs)
and the absence of LPM suppression, hadronic showers develop more
rapidly than electromagnetic showers with similar energies.  Here, we
use a simple model which generously overestimates the penetration of
hadronic showers: we treat them as electromagnetic showers, with each
generation having only twice as many particles as the
previous one.  Each generation develops over a distance $\Lambda$, the
hadronic interaction length; $\Lambda=83$ cm in water.  This
parameterization is shown by the dotted line in Fig. \ref{fig:length}.
For the energies for which data is available\cite{wigman}, this
thickness is more than enough for 99\% containment of hadronic
showers.  At high energies, this parameterization might overestimate
the shower length by 30-70\%.  Still, at $10^{20}$ eV, the shower is 3
times shorter than an electromagnetic shower; by $10^{23}$ eV, the
difference is a factor of 100.

In solids, high-energy $\pi^0$ (and some $\eta$) interact before they
can decay.  This happens when the decay length, $\gamma\beta c\tau$ is
larger than $\Lambda$.  In water, $\gamma\beta c\tau > \Lambda$ when
$E_\pi> 5$ PeV.  For the $\eta$, interactions predominate for
$E_\eta>3$ EeV.  So, a $10^{20}$ eV hadronic shower develops through
several generations before a significant electromagnetic component
develops.  The electromagnetic particle energies will be low enough
that the LPM effect will be absent.  Simulations confirm that the LPM
effect only occasionally affects hadronic showers, when a high energy
$\pi^0$ or $\eta$ decays\cite{zashadron}.

The dashed brown line in Fig. \ref{fig:length} shows the typical
length of hybrid $\nu_e$ showers that develop as shown in Fig
\ref{fig:schema}.  In F.g \ref{fig:length}, electrons take 80\% of the
neutrino energy and photons take 90\% of the electron energies.  As
Fig. \ref{fig:bremscaling} shows, this is a reasonable energy
partition.

Hybrid showers are much longer than purely hadronic showers because of
the length of the electron and photon tracks.  Above an energy of
$10^{20}$ eV, hybrid showers are much shorter than purely
electromagnetic showers.  The presence of photonuclear reactions
greatly shortens the $\nu_e$ showers.

Much of the length comes from the initial electron trajectory.  In
both treatments, electron energy loss due to direct pair production
and electronuclear interactions is neglect.  With a realistic
treatment of these effects, both electromagnetic and hybrid showers
would become shorter, and the fractional difference would increase.

Above $10^{23}$ eV, the hybrid shower length exceeds 1 km, comparable
to the typical thickness of the ice or water used for neutrino
detection.  Even if a vertically downward-going neutrino interacts
near the target surface, the bulk of the signal (which comes from the
end of the shower, when the number of particles is largest) will be
induced in the rock underneath the sensitive medium, limiting the
sensitivity to near-vertical showers; this may be particularly
relevant for SAUND.  Above 1 PeV, the earth is opaque to neutrinos, so
there are no corresponding upward-going neutrinos.

Some analyses avoid the shower-length problem by considering only
radiation from hadronic cascades from struck nuclei\cite{forte}.
Analyses that include emission from the 2nd, photon-produced hadronic
cascades may find lower limits and/or energy thresholds.

\section{Shower Lateral Distributions}

The characteristic lateral spread of electromagnetic shower is given
by the Moliere radius, $r_M = X_0 E_s/E_c = 6.2 $ cm in
water\cite{pdg}.  For hadronic showers, there is no corresponding
simple formula for lateral distributions.
 
To compare the widths of electromagnetic and hadronic showers in
solids (liquids should be similar), we consider data from the CERN LAA
project\cite{calorimeter}.  The collaboration compared the lateral
distributions of 5- 150 GeV electromagnetic and hadronic showers in a
lead/scintillating fiber calorimeter.  The showers were produced by
electron and $\pi^-$ interactions respectively.  They modelled the
electromagnetic lateral energy density as
\begin{equation}
\frac{dE}{dA} = \frac{A}{r} \frac{1}{(r^2+B^2)^2}
\end{equation}
where $A$ is the size of the signal, and $B$ is the shower width.
They found $B\approx 2$ cm, almost independent of energy.  They
parameterized the lateral distribution of hadronic showers with 2
components:
\begin{equation}
\frac{dE}{dA} = \frac{B_1}{r}e^{-r/\lambda_1} + \frac{B_2}{r}
e^{-r^2/\lambda_2^2}.
\end{equation}
where $B_1$ and $B_2$ are the sizes of the two components and
$\lambda_1$ and $\lambda_2$ are the lateral spreads of the two
components.  Neither $\lambda_1$, $\lambda_2$ nor $B_2/B_1$ varied
significantly with energy.

\begin{figure}[tb]
\center{\epsfig{figure=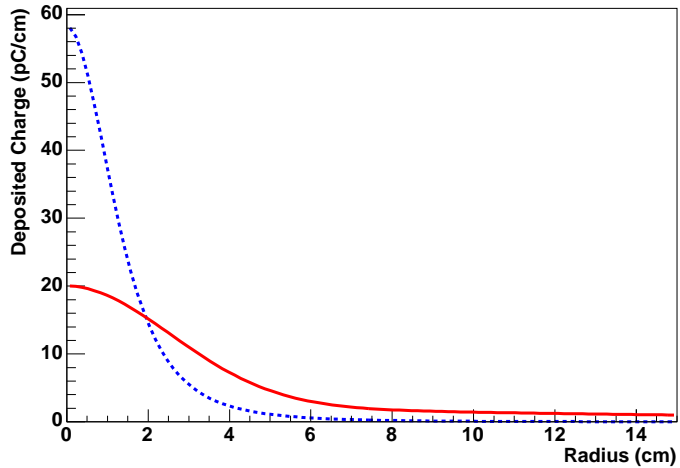,height=2.5in,clip=}}
\caption{Lateral profiles of charge deposition $dQ/dR$ for 150 GeV
electromagnetic (dashed line) and hadronic (solid line) showers as
observed by the LAA project.}
\label{fig:showercompare}
\end{figure}

Figure \ref{fig:showercompare} compares these energy depositions (in
terms of charge $Q$ deposited in the calorimeter as a function of
radius, $dQ/dr$), for 150 GeV electromagnetic and hadronic showers.
For electromagnetic showers, half of the energy is deposited within a
cylinder of 0.9 cm radius; for hadronic showers, 2.8 cm is required
for the same containment.  For 80\% containment radii of 1.9 cm and 11
cm are required for electromagnetic and hadronic showers,
respectively.  For 90\% containment, the radii are 2.8 cm and 21 cm.
Depending on the containment criteria, the hadronic showers have radii
3 to 8 times larger than electromagnetic showers.

The CERN LAA calorimeter showers developed largely in lead; water has
much lighter nuclei.  However, the lateral spread of the shower
depends on the transverse momentum of the particles produced in the
interactions; the $p_T$ distributions should be very similar for lead
and water.

In very high energies hadronic interactions, hard parton-parton
interactions will dominate the cross section. Scattered high
transverse momentum ($p_T$) partons will fragment into high $p_T$
hadrons\cite{who}.  At high collision energies, these high $p_T$
hadrons will widen the hadronic showers.  We do not estimate the
magnitude of the increased broadening here, but it could be
substantial.

In contrast, the transverse momentum of electromagnetic showers comes
largely from multiple scattering. The $p_T$ from multiple scattering
is independent of energy. A small fraction of the $p_T$ does come from
the pair production and bremsstrahlung reactions themselves.  When LPM
suppression is large, the mean opening angle (and, hence, $p_T$) in
bremsstrahlung and pair production increase by a factor $S$, the
suppression factor for the angle-independent
calculation\cite{lpmreview,gg}.  However, because multiple scattering
contributes most of the $p_T$, the shower width is not significantly
affected.

Because of the hadronic broadening with increasing collision energy,
the width ratios measured at 150 GeV should be treated as lower bounds
at $10^{20}$ eV.  Detailed simulations are needed to develop a better
estimate.  Even though the low-energy part of the shower is largely
electromagnetic, it retains the lateral spread acquired during it's
high-energy hadronic evolution; the CERN LAA measurements were
sensitive to both the hadronic and electromagnetic components of the
showers.

\section{Electromagnetic Radiation from Showers}

Coherent radio Cherenkov emission is dominated by the low-energy part
of the shower, where the net charge excess of $e^-$ over $e^+$ leads
to significant radiation\cite{slacexpt}.  The frequency spectrum of
the radiation depends on the width of the shower. An accurate
calculation of the frequency spectrum requires a full simulation and a
detailed calculation.  Here, we will consider some simple models which
should qualitatively illustrate the features which may be expected
from a full calculation.

Coherent Cherenkov radio emission occurs when the radio wavelength is
larger than the shower width; if the wavelength is small compared to
the width, coherence is lost. Alvarez-Muniz and Zas studied the effect
of lateral spreading by comparing the radiation from a full
3-dimensional electromagnetic cascade calculation of a 10 TeV shower
with a simplified 1-dimensional calculation\cite{zascalc}.  The
importance of the lateral spread depends on the observation angle.
Coherence is maximal when the observer looks along the Cherenkov
angle, $\theta_C=56^0$ in ice (and a similar value in the lunar
regolith\cite{glue}).  For frequencies above $\approx 500$ MHz,
Alvarez-Muniz and Zas found that the lateral spread significantly
reduces the radiation.  This is within the frequency range explored by
most of current experiments; these experiments are sensitive to the
lateral spreading.

An observer looking at the Cherenkov angle $\theta_C$ sees an apparent
shower thickness given by the lateral spread times $1/\sin(\theta_C)$
(about 1.21 in water).  In hybrid showers, the apparent thickness
may be several times higher than in purely electromagnetic showers; a
the lateral spread due to the hadronic interactions could have an
important effect at frequencies as low as 60-200 MHz. At higher
frequencies, there will be a significant loss of coherence, and the
radiation will be reduced.

Buniy and Ralston\cite{ralston} found that the radiation from a shower
can be found using the Fourier transform of the charge distribution.
They worked in the low frequency limit, where the lateral spread of
the form factor had no effect.  This approximation fails at high
frequencies.  The region of validity is much smaller for hybrid
showers than for purely electromagnetic ones.

The acoustic radiation from a shower comes from what is effectively a
line source which radiates largely perpendicular to the shower
direction; the shower looks like an expanding pancake.  The acoustic
frequency spectrum depends on the lateral spread of the shower.  For a
speed of sound of 1500 m/s\cite{acoustic}, the maximum frequency for
full coherence\ (i.e. the shower is contained within one acoustic
wavelength) drops from 75 kHz to 12.5 kHz when going from
electromagnetic to hybrid showers (for 80\% lateral containment).
This covers most of the frequency range studied by SAUND.

\section{Other Implications}

Although photonuclear interactions only dominate above $10^{20}$ eV,
they may affect showers at considerably lower energies, by introducing
hadronic components into largely electromagnetic showers.  This could
affect searches for $10^{18}$ eV neutrinos produced by cosmic ray
interactions with the cosmic microwave background\cite{halzen}.

At intermediate energies ($10^{16}$ - $10^{20}$ eV), photonuclear
interactions convert some electromagnetic shower energy into hadrons.
For example, a $10^{19}$ eV shower evolves through stages that include
$\approx 10$ $10^{18}$ eV particles and $\approx 100$ $10^{17}$ eV
particles, half photons and half electrons.  At $10^{18}$ and
$10^{17}$ eV, the probabilities for photonuclear interactions are
about 12\% and 3\% respectively.  These probabilities are high enough
that most showers with energies above $2\times 10^{18}$ eV will have
some hadronic component.  The produced hadrons may decay and introduce
a muon content into the shower. Although most $\pi^\pm$ and kaons
interact before they can decay, charm and bottom hadrons produced in
the shower may decay semi-leptonically, producing muons.  These muon
`tails' could be used to provide $\nu_e$ directional information in
experiments like IceCube\cite{icecube}.

The initial hadronic shower from the $\nu_e$ interaction, plus the
delayed hadronic shower after the electron to photon to hadronic
shower conversion could mimic a $\nu_\tau$ `double-bang'
event\cite{learned}.  The electron plus photon range equals
$\gamma\beta c\tau$ for a $\tau$ ($c\tau=290$ fs) at a neutrino energy
around 1 PeV.  The probability of photonuclear interactions at this
energy is small.  However, there are large fluctuations in $\tau$
decay length, in energy division (between the lepton and the target
nucleon) and measurement, and in shower development; all of these may
increase the likelihood of misreconstruction and consequent
misidentification.  An accurate estimate of the misidentification
probability requires detailed simulations. 

\section{Uncertainties}

There are significant uncertainties in these calculations.   There are
approximations in Migdals calculations and uncertainties in the
additional suppression mechanisms and the photonuclear calculations.

Migdals calculations do a good job of describing the SLAC E-146 and
CERN data, with the apparent exception of the E-146 data on carbon
targets\cite{e146prl}.  Nevertheless, they have some limitations.
Migdal assumed that the scattering was Gaussian; A Gaussian
distribution considerably underestimates the number of large-angle
Coulomb scatters, and could therefore, under-estimate the suppression.
Migdal neglected interactions with the atomic electrons in the target.
This may be important for low-$Z$ targets.

Two newer and more sophisticated calculations, by
Zakharov\cite{zakharov} and by Baier and Katkov\cite{bk}, remedy these
problems.  Both include accurate models of Coulomb scattering and
account for atomic electrons by using separate elastic and inelastic
potentials.  Both calculations match the experimental data.
Unfortunately, code for these calculations is not publicly available
for use in simulations.  However, Baier and Katkov give a cross
section for high-energy ($k\gg E_{LPM}$) pair conversion which agrees
with Migdals calculation to within about 20\%, well within the
accuracy needed here.

None of these calculations explicitly consider hydrogen targets.
Hydrogen is problematic because the standard Thomas-Fermi screening
calculations are only accurate for atomic numbers $Z>5$.  A
hydrogen-specific screening correction is required to accurately find
the cross sections\cite{tsai}.  However, because pair production in
water is dominated by interactions with oxygen, the error in the
hydrogen screening causes much less than a 10\% effect on the cross
sections for water.

Other suppression mechanisms may enter at very high energies.  When
the formation length (reaction zone) is long enough, a nascent photon
may interact, by either pair production or photonuclear interactions)
before it is fully formed.  Bremsstrahlung and pair production may
suppress each other, and photonuclear interactions may suppress
bremsstrahlung.  These effects may be important when the formation
length (including the LPM effect) is larger than $X_0$ or
$1/\sigma\rho$.  The former can only occur when $E>E_p$, where $E_p =
540$ TeV in water\cite{lpmreview}.  At this energy, it only applies
for a very narrow range of $k/E$ (or, for photons, $E/k$).  The range
of $k/E$ (or $E/k$) where it applies is only significant at much
higher energies; above $10^{20}$ eV.  However, this mechanism won't
change the conclusion that photonuclear reactions dominate above
$10^{20}$ eV.  If anything, the mutual suppression further reduces the
electromagnetic cross sections, strengthening this conclusion.

A broader question for all of the electromagnetic interactions
involves higher-order reactions like $eN\rightarrow e^+e^- eN$ (direct
pair production) and $\gamma N\rightarrow
e^+e^-\gamma$\cite{lpmreview}.  Normally, the cross section for these
higher order reactions are a factor of order $\alpha = 1/137$ smaller
than the leading-order processes.  However, the higher-order processes
require a higher momentum transfer from the target, and so are much
less subject to LPM suppression.  When LPM suppression is large, these
higher-order reactions could dominate, creating a floor for the
electromagnetic cross sections.  Detailed comparisons have not been
done, but $1/137$ suppression is only reached at energies above
$10^{20}$ eV.

The final caveat involves photonuclear interactions.  The
extrapolation to $10^{20}$ eV is a factor of 2000 in $W$ beyond the
HERA data.  Between the possible moderation of the $W^{0.16}$ Pomeron
trajectory and uncertainties in direct $\gamma q$ cross sections, the
uncertainties are considerable.  However, because of the slow energy
variation, the large extrapolation uncertainty cannot radically change
the crossover energy.  Even a 50\% reduction in the rise in cross
section above the HERA data (i.e. for $k>20$ TeV) would only increase
the crossover energy by about 40\%.

In short, although there are significant uncertainties present, even
generous error estimates do not affect the conclusion that
photonuclear cross sections are larger than electromagnetic ones for
$\nu_e$ energies above $10^{20}$ eV.

\section{Conclusions}

At energies above $10^{20}$ eV, photons are more likely to interact
hadronically than through pair production, and $\nu_e$ showers are
likely to be hadronic. By $10^{21}$ eV, hadronic showers are 4
times as frequent as electromagnetic showers. These hadronic showers
are considerably shorter, and several times wider than purely
electromagnetic showers.  The increased width of the showers reduces
the frequencies at which radio and acoustic radiation are coherently
emitted, and may affect the conclusions of experiments that study
radio and/or acoustic emission from $\nu_e$ showers.

It is a pleasure to acknowledge useful discussions about photonuclear
interactions with Ralph Engel and Mark Strikman.  Dmitri Chirkin and
Bob Stokstad made useful comments about this manuscript.  It is also a
pleasure to thank the UCB and LBNL IceCube groups for many useful
pointers.  I also thank the workshop organizers for putting together a
very interesting week.  This work was funded by the U.S. National
Science Foundation under Grant number OPP-0236449 and the
U.S. Department of Energy under contract number DE-AC-76SF00098.

\end{document}